\documentclass[10pt,prx,aps,twocolumn,showpacs,amsmath,amssymb,superscriptaddress,floatfix]{revtex4-2}

\usepackage{graphicx}
\usepackage[export]{adjustbox}
\usepackage{dcolumn}
\usepackage{upgreek}
\usepackage{bm,dsfont}
\usepackage{amssymb,mathtools}
\usepackage{booktabs}
\usepackage{color}
\usepackage{microtype}
\usepackage[T1]{fontenc}
\usepackage{lmodern}
\usepackage[shortlabels]{enumitem}
\usepackage{physics}
\usepackage{derivative}
\usepackage{tikz}
\usepackage{tikz-cd}
\usepackage{placeins}
\usepackage[colorlinks,plainpages=false,pdfpagemode=UseNone,pdfstartview=FitBH]{hyperref}
\usepackage{sidecap}
\allowdisplaybreaks
\hypersetup{
    colorlinks = true,
    linkcolor = [rgb]{0.87,0.18,0.22},
    citecolor = [rgb]{0.0,0.60,0.32},
    urlcolor = [rgb]{0.20, 0.30, 0.54}}
    \definecolor{eggplant}{RGB}{180,33,147}
    \makeatletter\renewcommand{\fnum@table}[1]{\tablename~\thetable.}\makeatother

\definecolor{citecolor}{rgb}{0.0,0.60,0.32}

\graphicspath{{./figs}}
\begin{document}

\title{Neural network impurity solver for real-frequency dynamical mean-field theory}

\author{Fenglin Deng}
\affiliation{Suzhou Institute for Advanced Research, University of Science and Technology of China, Suzhou 215123, China}
\affiliation{School of Artificial Intelligence and Data Science, University of Science and Technology of China, Suzhou 215123, China}

\author{Yi Lu}
\affiliation{National Laboratory of Solid State Microstructures and Department of Physics, Nanjing University, Nanjing 210093, China}
\affiliation{Collaborative Innovation Center of Advanced Microstructures, Nanjing University, Nanjing 210093, China}

\author{Xiaodong Cao}
\email{xdcao@ustc.edu.cn}
\affiliation{Suzhou Institute for Advanced Research, University of Science and Technology of China, Suzhou 215123, China}
\affiliation{School of Artificial Intelligence and Data Science, University of Science and Technology of China, Suzhou 215123, China}

\author{Zhicheng Zhong}
\affiliation{Suzhou Institute for Advanced Research, University of Science and Technology of China, Suzhou 215123, China}
\affiliation{School of Artificial Intelligence and Data Science, University of Science and Technology of China, Suzhou 215123, China}

\date{\today}
\begin{abstract}
We introduce a neural network impurity solver for real-frequency DMFT that employs a multihead cross-attention mechanism to map hybridization functions to spectral functions, conditioned on impurity parameters.
Trained on high-quality MPS data from complex contour time evolution and incorporating derivative constraints with respect to the complex-time angle, our model achieves smooth generalization to the real-frequency axis.
Benchmarking on the single-band Hubbard model for the Bethe lattice demonstrates quantitative accuracy across metallic, strongly correlated, and insulating regimes.
\end{abstract}

\maketitle

\section{Introduction}
Developing accurate and efficient computational methods for strongly correlated systems is one of the central topics in condensed matter physics. Due to coupling across a wide range of energy scales, these systems require non-perturbative analysis such as dynamical mean-field theory (DMFT)~\cite{Georges1996DMFTReview,Kotliar2006DMFTReview}. 
It has been shown that DMFT is capable of capturing a variety of correlation-driven phenomena and, when combined with first-principles methods, now serves as a standard framework for investigating strongly correlated materials beyond the single-particle description.~\cite{Rozenberg1994MIT,Georges2013HundReview,Veenstra2014SOC,Shinaoka2015SOC}. 

A key step in DMFT is the solution of a quantum impurity model at each iteration of the self-consistency loop. Although this impurity problem is simpler than the original lattice model, it remains highly nontrivial due to its intrinsic quantum many-body nature. A variety of impurity solvers have been developed so far, including exact diagonalization (ED)~\cite{Caffarel1994ED}, continuous-time quantum Monte Carlo (CT-QMC)~\cite{Gull2011CTQMCReview}, numerical renormalization group (NRG)~\cite{wilson1975renormalization,Bulla2008NRGReview}, and tensor-network approaches~\cite{White1992DMRG,White1993DMRG,Schollwock2011DMRGReview,mpsdmft2015,cao2021tree,cao2024ftmetts,Daniel_fork}.
Each of these many-body solvers has its own strengths and limitations, is typically reliable only in certain parameter regimes, and usually entails significant computational costs. In addition, conventional solvers treat each set of impurity parameters as an entirely new problem, with little or no reuse of information from previously solved instances. Taken together, these factors severely constrain studies that require a large number of impurity solves—for example, large-scale real-space DMFT for spatially inhomogeneous systems~\cite{Nolting_rDMFT,Freericks_rDMFT} or schemes that replace a large cluster by many smaller impurity problems~\cite{Imada_rrDMFT}. These challenges motivate the development of a fast, accurate impurity solver that can efficiently transfer knowledge across the impurity parameter space.

In this vein, data-driven machine learning (ML) impurity solvers have recently been explored~\cite{Arsenault2014MLAIM,Arsenault2015MLDMFT,rogers2021,Weber2021,agapov2024,kakizawa2024pinn,Lee2019a,Lee2019b,sturm2021,Ren_2021}. 
Trained on datasets generated by established quantum many-body solvers, these methods can directly predict impurity Green’s functions across broad parameter regimes, thereby enabling fast evaluation of desired physical observables such as the Green's function.
A reliable machine learning impurity solver relies on two pillars. First, the training data must be of high quality. Second, the solver must remain accurate and stable under repeated updates of the impurity parameters to ensure DMFT convergence, which in turn requires that the neural network generalizes well across the impurity parameter space.

Most existing neural network solvers are demonstrated only as one-shot predictors for the impurity model without being embedded in a full DMFT self-consistency loop, especially on the real-frequency axis. This is mainly due to the following challenges associated with learning directly in real frequency.
On the data side, high quality real-frequency data are generally difficult to generate.
CT-QMC yields imaginary-time data contaminated by Monte Carlo noise and requires ill-conditioned analytic continuation to access real-frequency observables; ED is restricted to only a few bath sites (typically $\sim 10$), resulting in a poor representation of the continuum; NRG is accurate at low energies but less reliable for high energy fine structure; tensor-network solvers suffer from rapid bond-dimension growth during long-time evolution, hampering access to low-frequency physics.
On the machine learning side,
real-frequency spectra often contain sharp, fine-scale features that are difficult to learn. As a result, it has remained unclear whether an neural network impurity solver can robustly close the DMFT loop directly on the real-frequency axis.

In this manuscript, we present a neural network impurity solver that enables fast and accurate predictions of the real-frequency spectral function from the impurity parameters. The training data are generated by a complex-time tensor-network impurity solver~\cite{Cao2024a}, which evolves along a contour with a finite complex-time angle, thereby substantially reducing the computational cost compared to real-time evolution. 
The solver is trained on finite-angle contours and is regularized by a loss term on the output's first derivative, which ensures stable extrapolation and yields accurate spectral functions directly on the real-frequency axis.
We further design an architecture that efficiently captures real-frequency spectral features and fuses information from the hybridization function and interaction parameters via cross-attention~\cite{xu2016showattendtellneural,bahdanau2016neuralmachinetranslationjointly,Vaswani2017Attention,dosovitskiy2021imageworth16x16words}. Together, these improvements yield a practical real-frequency solver that robustly closes the DMFT self-consistency loop across a broad parameter regime.

The paper is organized as follows. Section~\ref{sec::method::cpt_solver} introduces the complex-time solver used to generate the training data. Section~\ref{sec::method::architecture} describes the network architecture. The loss function and computational details are given in Section.~\ref{sec::method::loss_function} and Appendix~\ref{sec::method::comp_details}. 
Section ~\ref{sec::results} provides benchmark results on the single-band Hubbard model and discusses the learned latent-space representation.
Section~\ref{sec::conclusion} concludes with an outlook.

\section{Method}\label{sec::method}
This section presents the general framework of our neural network impurity solver developed for real-frequency DMFT studies of the single-band Hubbard model on the Bethe lattice.
We first introduce the complex-time impurity solver used to generate the training data, and then elaborate on the neural network architecture and training details.

\subsection{Complex-time solver} \label{sec::method::cpt_solver}
The DMFT framework maps the lattice problem to a single-impurity Anderson model (SIAM) with Hamiltonian
\begin{align}
    \hat{H} = &\sum_\sigma \epsilon_d\, \hat{d}^\dagger_\sigma \hat{d}_\sigma + U\, \hat{n}_\uparrow \hat{n}_\downarrow \nonumber\\
    &+ \sum_{k\sigma} \epsilon_k\, \hat{c}^\dagger_{k\sigma} \hat{c}_{k\sigma}
    + \sum_{k\sigma} \left( V_k\, \hat{c}^\dagger_{k\sigma} \hat{d}_\sigma + {\rm h.c.} \right),
\end{align}
where $\hat{d}_\sigma$ ($\hat{d}^\dagger_\sigma$) annihilates (creates) an electron on the impurity with spin $\sigma\in\{\uparrow,\downarrow\}$, $\hat{n}_\sigma=\hat{d}_\sigma^{\dagger} \hat{d}_\sigma$ is the density operator, $k$ is the bath state index, $U$ is the on-site Hubbard interaction, and $\epsilon_d$ and $\epsilon_k$ are the impurity and bath energy levels. $V_k$ characterizes the hybridization between the impurity and the bath, which is obtained by discretizing the hybridization function $\Delta(\omega)$ into $N_b$ equal intervals $\{I_k\}$
\begin{align}
  	V_k=\int_{I_k} \left[-\frac{1}{\pi}\mathrm{Im}\Delta(\omega)\right]d\omega, \\ \nonumber
  	\epsilon_k = \frac{1}{V_k}\int_{I_k}\omega \left[-\frac{1}{\pi}\mathrm{Im}\Delta(\omega)\right]d\omega.
\end{align}
Since our system preserves the spin SU$(2)$ symmetry, we omit the spin index in the following discussion. 

\begin{figure*}[t]
    \centering
    \includegraphics[width=2.0\columnwidth]{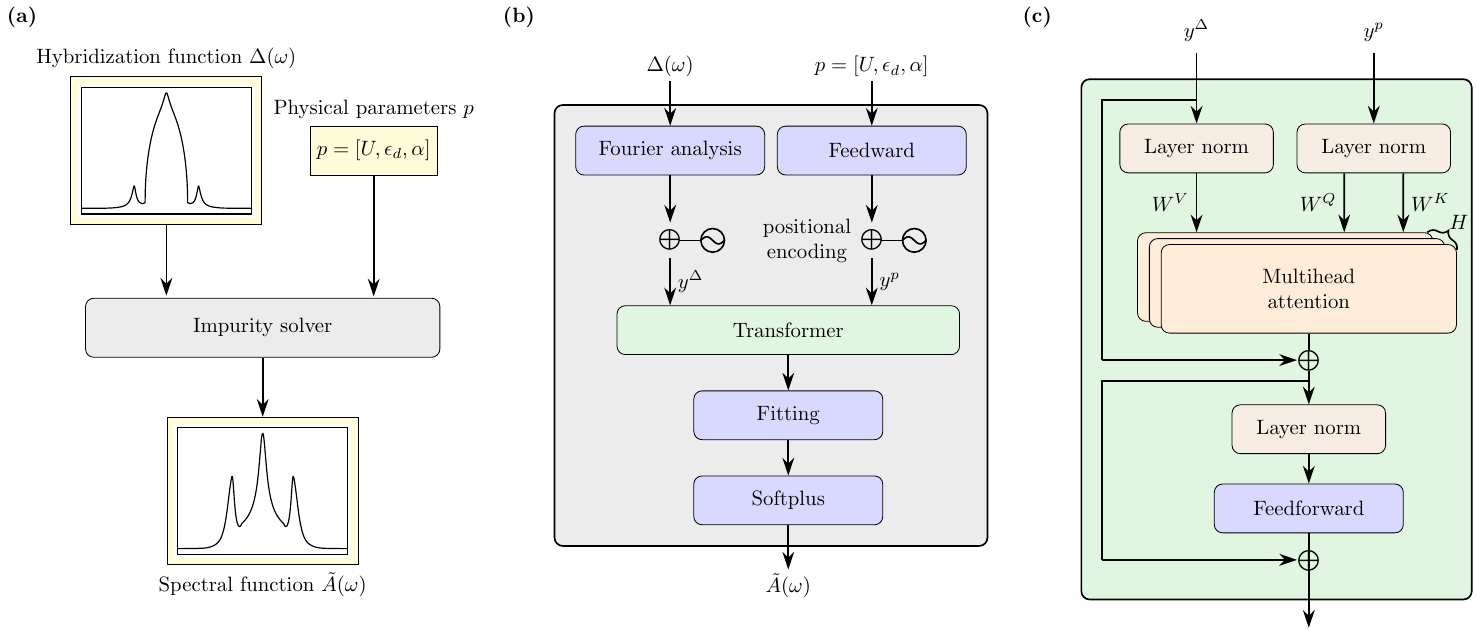}
    \caption{
    Schematic illustration of the neural network impurity solver. (a) The impurity solver takes as input the hybridization function $\Delta(\omega)$ together with the physical parameters $p=[U,\epsilon_d,\alpha]$, and produces the corresponding spectral function $\tilde{A}(\omega)$. (b) Neural network architecture of the solver. The hybridization function and the physical parameters are processed separately through FAN layers and feedforward networks, respectively. Positional encodings are added to these representations, which are then passed through a transformer block, followed by fitting layers and a softplus projection to generate the predicted spectral function $\tilde{A}(\omega)$. (c) Internal structure of a transformer block. The hybridization function and parameter representations are first normalized (Layer norm) and then coupled through multihead attention. The attention output is followed by a second Layer norm and a feedforward network, each equipped with residual connections. 
    }
    \label{fig:nn_architecture}
\end{figure*}

To achieve the self-consistent condition that the impurity Green's function is equal to the local lattice one, we iteratively update the impurity Green's function and the hybridization function until convergence is reached. 
During each real-frequency DMFT iteration, we obtain the impurity retarded Green’s function on the real-frequency axis $G^R(\omega)$ from an impurity solver, and then update the hybridization function
\begin{align}
    \Delta(\omega) = \frac{D^2}{4} A(\omega),
\end{align}
where $A(\omega)=-\frac{1}{\pi}\Im G^R(\omega)$, and $D$ is the half-bandwidth of the initial semielliptic spectrum $-\frac{1}{\pi}\mathrm{Im}\Delta(\omega)=\frac{2}{\pi D}\sqrt{1-\left( \frac{\omega}{D}\right)^2 }$. 

In this work, we compute the retarded Green's function $G^R(\omega)$ using a complex-time impurity solver~\cite{Cao2024a} in the natural orbital basis ~\cite{Lu2014,Lu2019,Lu2021}  rather than the usually employed real-time approach~\cite{mpsdmft2015}. 
We define a generalized retarded Green's function $G^{R}(t, \alpha)$ on the complex-time contour
\begin{align}\label{eq::retard_gt_complex}
G^{R}(t, \alpha) = \theta(t) [G^{+}(t, \alpha)-G^{+}(t, \alpha)],
\end{align}
where $G^{+}$ ($G^{-}$) is the greater (lesser) Green's function
\begin{align}\label{eq:Ggreater}
    G^{+}(t, \alpha)
    &= -\mathrm{i}\langle \psi_g | \hat{d} \ket{\psi^{+}(t,\alpha)} , \\ \nonumber
    G^{-}(t, \alpha)
    &= \mathrm{i}\langle \psi_g | \hat{d}^\dagger \ket{\psi^{-}(t,\alpha)},
\end{align} 
with $\ket{\psi^{\pm}(t, \alpha)}$ the excited states evolved along a complex-time contour $z^{\pm}(t,\alpha) = \mathrm{e}^{\pm \mathrm{i}\alpha} t$
\begin{align}
    \ket{\psi^{+}(t,\alpha)} = e^{- \mathrm{i} (\hat{H}-E_g) z^{+}(t,\alpha) } \hat{d}^\dagger | \psi_g \rangle, \\ \nonumber
    \ket{\psi^{-}(t,\alpha)} = e^{ \mathrm{i} (\hat{H}-E_g) z^{-}(t,\alpha) } \hat{d} | \psi_g \rangle.
\end{align}
Within our impurity solver, the ground state $|\psi_g\rangle$ with energy $E_g$ is represented as a matrix product state (MPS), optimized by the density matrix renormalization group (DMRG) \cite{White1992DMRG,White1993DMRG,Schollwock2011DMRGReview}.  The subsequent time evolution is performed using the time-dependent variational principle (TDVP) \cite{Haegeman2011TDVP,Haegeman2013PostMPS,Haegeman2016TDVPMPS,Vanderstraeten2019TangentMPS}. 
By employing a complex-time contour with a finite imaginary component ($\pm \Im z^{\pm}(t,\alpha) < 0$), the evolution of excited states $\ket{\psi^{\pm}(t,\alpha)}$ incorporates a damping factor that suppresses high-energy excitations. This suppression enables accurate long-time propagation at a substantially reduced computational cost, as it enables a much smaller bond dimension $\chi$ than that for real-time evolution.

The real-time Green's function is given by the $\alpha=0$ limit in Eq.~\ref{eq::retard_gt_complex}, and can be reconstructed for small $\alpha$ via a Taylor expansion, as detailed in Ref.~\onlinecite{Cao2024a}.
In this work, we use a first-order approximation
\begin{align}\label{eq:complexG}
G^{R}(t, \alpha=0) \approx G^{R}(t, \alpha)-\alpha \partial_\alpha G^{R}(t, \alpha),
\end{align}
where the first order derivative term is given by
\begin{align}
\partial_\alpha G^{R}(t, \alpha) &= \theta(t) [\partial_\alpha G^{+}(t, \alpha)-\partial_\alpha G^{-}(t, \alpha)],
\\ \nonumber
\partial_\alpha G^{+}(t, \alpha) &= -\mathrm{i}z^{+} \langle \psi_g | \hat{d} (\hat{H}-E_g) |\psi^{+}(t,\alpha)\rangle,\\ \nonumber
\partial_\alpha G^{-}(t, \alpha) &= \mathrm{i}z^{-} \langle \psi_g | \hat{d}^\dagger (\hat{H}-E_g) |\psi^{-}(t, \alpha)\rangle.
\end{align}
Computing the first-derivative term requires minimal additional cost because the most expensive component—the time-evolved state $\ket{\psi^\pm(t,\alpha)}$ has already been obtained when computing $G^R(t,\alpha)$.
The approximated real-frequency spectral function is then given by
\begin{align}
A(\omega) &= -\frac{1}{\pi} {\rm Im} \int_0^\infty \mathrm{d}t\, e^{\mathrm{i}\omega t} G^{R}(t) \nonumber \\
&\approx A(\omega, \alpha) - \alpha \partial_\alpha A(\omega, \alpha) \coloneqq \tilde{A}(\omega, \alpha),
\end{align}
where $A(\omega, \alpha)$ and $\partial_\alpha A(\omega, \alpha)$ are the Fourier transforms of $G^{R}(t, \alpha)$ and $\partial_\alpha G^{R}(t, \alpha)$. 
This approximation is exact at $\alpha = 0$ ($\tilde{A}(\omega) = A(\omega)$), with the deviation increasing as $\alpha$ increases. Therefore, we use relatively small complex angles ($\alpha \approx 0.1$) in our training dataset (see Sec.~\ref{sec::results} for details), where the first derivative restores the low-frequency spectra well, while higher-order terms in the expansion in $\alpha$ mainly contribute to fine structures in the high-energy part~\cite{Cao2024a}.

\subsection{Network structure} \label{sec::method::architecture}

The neural network we design predicts the real-frequency impurity spectral function $\tilde{A}(\omega)$ from the hybridization function $\Delta(\omega)$ and physical parameters $\mathbf{p} = [U, \epsilon_d, \alpha]$ directly. The model comprises three blocks: (i) embedding of the physical parameters and hybridization function, (ii) cross-attention fusion, and (iii) spectral projection.
The full architecture is illustrated in Fig.~\ref{fig:nn_architecture}, and we detail each block of the model below.

\paragraph*{Embedding of physical parameters and hybridization function.}
We use a multi-layer feedforward network (FFN) to embed the physical parameters $\mathbf{p}=[U, \epsilon_d, \alpha]$ as
\begin{align}
\mathbf{p}^{(i+1)}
  = \mathrm{Swish}( W_i \mathbf{p}^{(i)} + b_i ),
  \qquad i = 0,\dots,L-1,
\end{align}
where $\mathbf{p}^{(0)} = \mathbf{p}$, $L$ is the number of layers, and $W_i$, $b_i$ are learnable weights and biases. 
This embedding procedure transforms the physical parameters $\mathbf{p}$ into $\mathbf{p}_\mathrm{emb} \in \mathbb{R}^{d_\mathrm{emb}}$.
A learnable positional encoding is added afterward to distinguish the physical parameters and preserve their relative order in the input vector.

The hybridization function $\Delta(\omega) \in \mathbb{R}^{N_\omega}$ is embedded using a stack of Fourier Analysis Network (FAN) layers~\cite{dong2025fan}. A single FAN layer applies two independent linear projections
\begin{align}
\mathbf{q} &= W_q \, \Delta(\omega) + b_q, \\
\mathbf{g} &= W_g \, \Delta(\omega) + b_g,
\end{align}
and produces the feature vector by applying different nonlinear activation functions
\begin{align}
 \mathbf{f} = \big[ \sin(\mathbf{q}), \;\cos(\mathbf{q}),\; \mathrm{GELU}(\mathbf{g}) \big].
\end{align}
By stacking multiple FAN layers, the hybridization function $\Delta(\omega)$ is transformed into $\Delta_\mathrm{emb} \in \mathbb{R}^{d_\mathrm{emb}}$. 
Similar to the parameter embedding, a learnable positional encoding is added afterward to maintain frequency ordering.

\paragraph*{Cross-attention between parameters and hybridization function.}
To fuse the physical parameter embedding $\mathbf{p}_\mathrm{emb}$ with the
hybridization embedding $\Delta_\mathrm{emb}$, we use a standard multihead cross-attention mechanism. 
For each head $h = 1,\dots,H$, the embeddings are projected to $d_h=d_{\mathrm{emb}}/H$
\begin{align}
\mathbf{q}_h = W_h^Q \mathbf{p}_\mathrm{emb}, \, \,
\mathbf{k}_h = W_h^K \mathbf{p}_\mathrm{emb}, \, \,
\mathbf{v}_h = W_h^V \Delta_\mathrm{emb}.
\end{align}
The head output is then computed via a weighted value projection with an attention score matrix
\begin{align}
\mathbf{o}_h = \alpha_h \mathbf{v}_h, \quad \alpha_h = \mathrm{softmax} \!\left( \frac{\mathbf{q}_h \mathbf{k}_h^\top}{\sqrt{d_h}} \right).
\end{align}
Each transformer block output is obtained by first concatenating the outputs from all heads and then linearly projecting them back to the original embedding dimension $d_\mathrm{emb}$
\begin{align}
\mathbf{h}_\mathrm{out} = W^O \big[\, \mathbf{o}_1;\dots;\mathbf{o}_H \,\big],
\end{align}
where $W^O \in \mathbb{R}^{d_\mathrm{emb} \times d_\mathrm{emb}}$, $\mathbf{h}_\mathrm{out} \in \mathbb{R}^{d_\mathrm{emb}}$, and further processed by a two layer FFN layer with $\text{GELU}(\cdot)$ activation.
See Fig.~\ref{fig:nn_architecture} for additional details.

\paragraph*{Output projection.}
The output layer converts the representation from the preceding transformer blocks into a physically meaningful spectral function $\tilde{A}(\omega) \in \mathbb{R}^{N_\omega}$ via a linear projection followed by a Softplus activation to enforce positivity
\begin{align}
\tilde{A}(\omega) = \mathrm{softplus} \Big( W_\mathrm{out}\, \mathbf{h}_\mathrm{out} + b_\mathrm{out} \Big).
\end{align}

\subsection{Loss function}\label{sec::method::loss_function}

The loss function we use to train the neural network contains a main loss function and an L2 weight regularization with coefficient $10^{-4}$
\begin{align}
    \mathcal{L}_{\mathrm{total}} &= 
        \mathcal{L}_{\mathrm{main}}
        + \mathcal{L}_{\mathrm{L2}} .
\end{align}
The main loss contains discrepancies in both the predicted spectral function and its first order derivative with respect to the complex-time angle $\alpha$
\begin{align}
    \mathcal{L}_{\mathrm{main}} &=
    \left\lVert \tilde{A}^{\mathrm{pred}}(\omega, \alpha) - \tilde{A}^{\mathrm{data}}(\omega, \alpha) \right\rVert \nonumber \\
    &+ \gamma \left\lVert
      \partial_{\alpha} \tilde{A}^{\mathrm{pred}}(\omega, \alpha) -
      \partial_{\alpha} \tilde{A}^{\mathrm{data}}(\omega, \alpha)
    \right\rVert , \label{eq:loss}
\end{align}
where $\gamma$ is a tunable parameter.
The $\alpha$-derivative term constrains the variation of the predicted spectral function along the $\alpha$ direction in parameter space, promoting a smooth, physically consistent mapping and better generalization.
See Sec.~\ref{sec::results} for further discussions regarding the derivative term, where we show that this term is crucial for reliable extrapolation to the real-frequency limit ($\alpha=0$) from a training set containing only finite-angle data.

\section{Results and discussion}\label{sec::results}
In this section, we evaluate our neural network impurity solver by reporting self-consistent DMFT results, including the spectral function, self-energy, and quasiparticle weight at half filling and doped regimes.
The datasets we use span $U \in (0,3.2D]$ and $\epsilon_d + 0.5U \in [-1.04D,1.04D]$, covering both metallic and insulating regimes. The training data consist of spectral functions $\tilde{A}(\omega, \alpha)$ at finite complex-time angles $\alpha \in [0.075,0.150]$,  
while testing is performed on the physical spectral functions $A(\omega)$ compared with the real-time MPS solver. 
In total, the dataset comprises 4,942 training samples and 499 test samples. 
We further analyze the intermediate outputs of the cross-attention blocks to elucidate the latent-space representations learned by the trained network.
Computational details—including solver parameters, frequency grids, network hyperparameters, and training settings—are provided in Appendix~\ref{sec::method::comp_details}.

\subsection{Half filling} \label{sec::results::half_filling}

\begin{figure}[t]
    \centering
    \includegraphics[width=1.0\columnwidth]{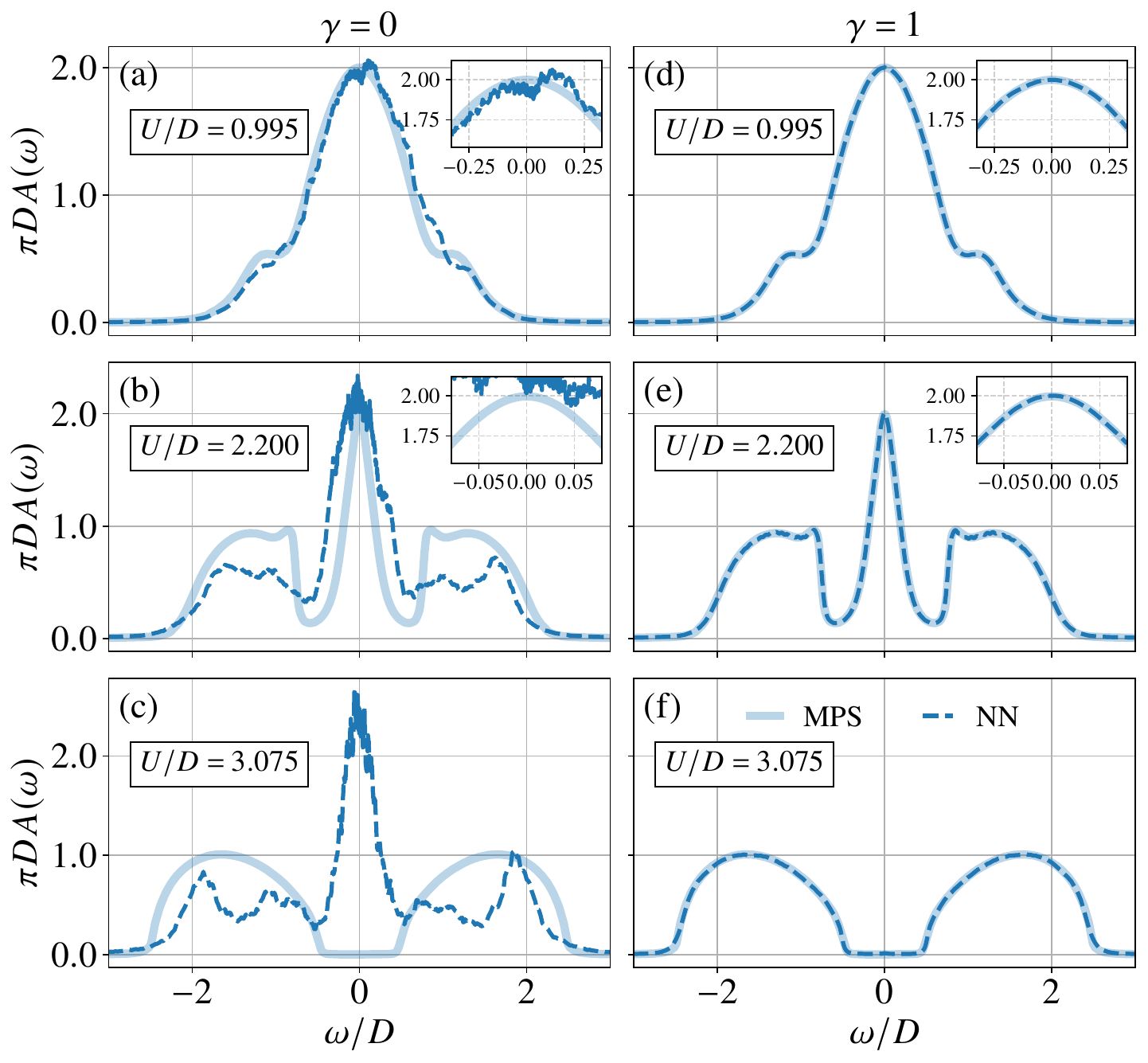}
    \caption{Spectral functions for different interaction strengths $U$. Panels (a)--(c) show results obtained without the $\alpha$-derivative constraint, while panels (d)--(f) correspond to results with the constraint. Thick transparent lines are reference results from the MPS real-time solver, and dashed lines denote predictions from the neural network solver. Insets show zoomed regions around $\omega=0$.}
    \label{fig:dmftnn_half_spectral}
\end{figure}

\begin{figure}[t]
    \centering
    \includegraphics[width=1.0\columnwidth]{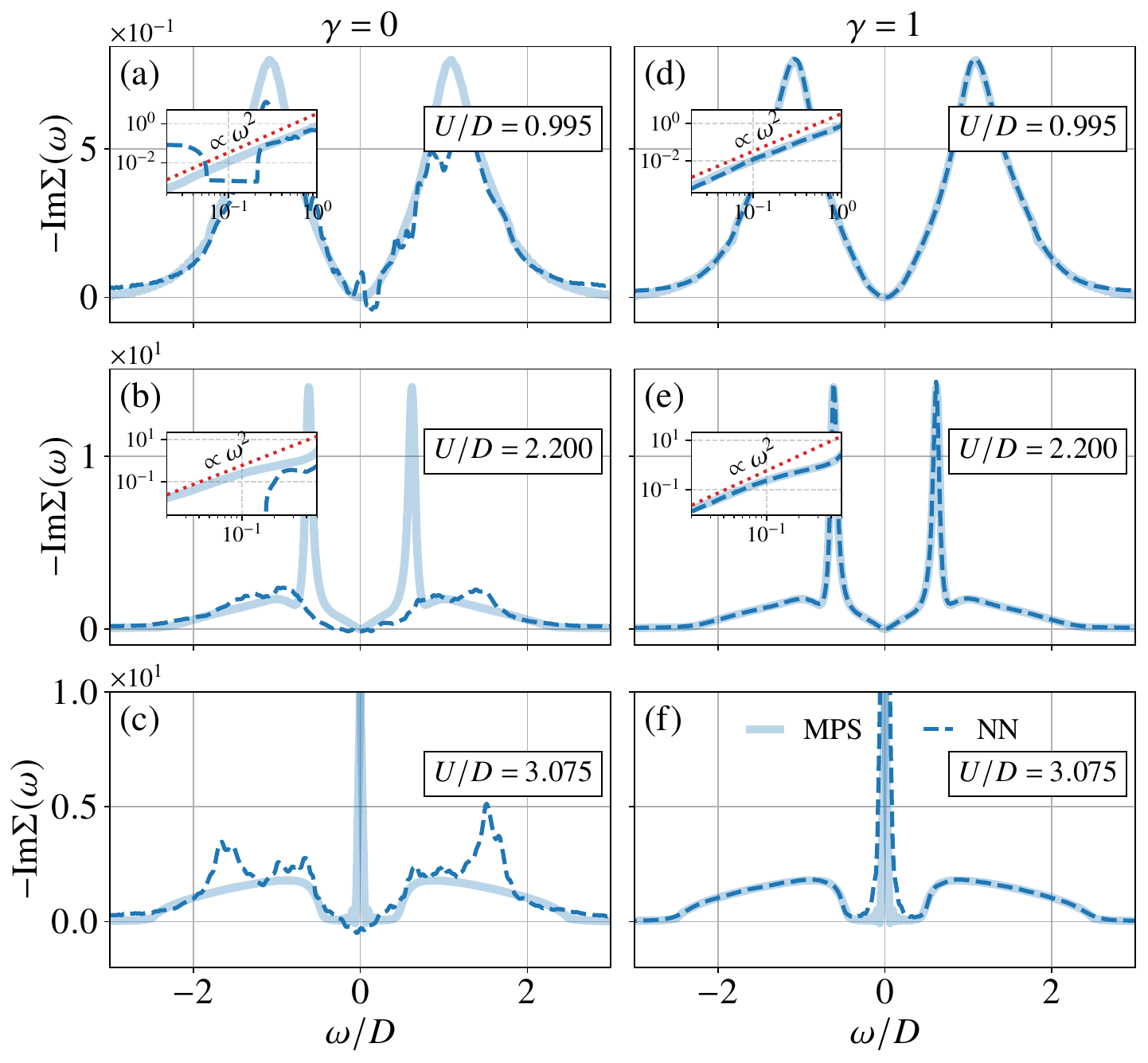}
    \caption{Self-energy for the same interaction strengths $U$ as in Fig.~\ref{fig:dmftnn_half_spectral}, showing results without( Panels (a)-(c) ), and with( Panels (d)-(e) ) the $\alpha$-derivative constraint. Thick lines represent real-time MPS reference results, and dashed lines show neural network predictions. Insets highlight the low-frequency region near $\omega=0$, and the red dotted line indicates a Fermi liquid behavior of $-\text{Im}\Sigma(\omega)\sim\omega^2$.}
    \label{fig:dmftnn_half_sigma}
\end{figure}

We first benchmark the neural network impurity solver by performing DMFT calculations at half filling.
We choose three representative interactions $U/D = 0.995$, $2.2$, and $3.075$, which correspond to weakly correlated, strongly correlated metallic, and Mott-insulating regimes.
Fig.~\ref{fig:dmftnn_half_spectral} presents the real-frequency spectral functions $A(\omega)$ given by two networks trained with $\gamma = 0$ and $\gamma = 1$.
It's clear from the figure that without the derivative constraint, the network fails to reproduce the overall spectral shape, whereas including the derivative term yields spectra that accurately match the reference DMFT results.
To further benchmark the accuracy of the solvers, we compare the corresponding self-energies in Fig.~\ref{fig:dmftnn_half_sigma}.
The results show that imposing the $\alpha$-derivative constraint produces predictions in close agreement with the reference, while omitting it leads to noticeable discrepancies.
These results unambiguously show that our neural network impurity solver is able to provide high-accuracy predictions, including the low energy Fermi liquid behavior $-\text{Im}\Sigma(\omega) \sim (\omega/D)^2$ (see insets of Fig.~\ref{fig:dmftnn_half_sigma}).
Moreover, the comparison between the two networks highlight the importance of incorporating physically motivated derivative information into the loss function in order to achieve stable extrapolation from finite complex angles $\alpha > 0$ to the real-frequency limit $\alpha = 0$.

\begin{figure}[t]
    \centering
    \includegraphics[width=1.00\columnwidth]{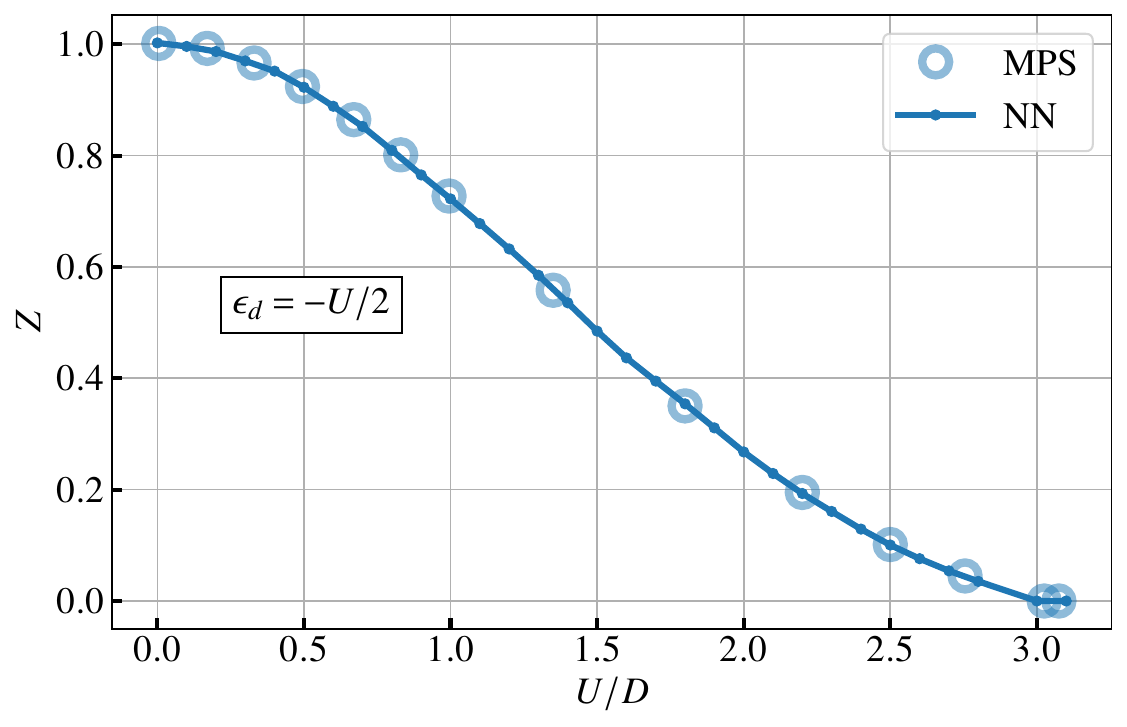}
    \caption{Quasiparticle weight predicted by the neural network solver compared to the reference.}
    \label{fig:qpw}
\end{figure}

To evaluate the performance of our neural network impurity solver across interaction strengths and its ability to capture low-energy physics, we benchmark the quasiparticle weight $Z=\bigl[ 1-\frac{\partial \text{Re}{\Sigma}(\omega)}{\partial\omega}|_{\omega=0} \bigr]^{-1}$ against reference data for $0 \leq U/D \leq 3.1$.
As shown in Fig.~\ref{fig:qpw}, the model accurately reproduces Fermi-liquid behavior in the metallic regime and successfully captures the metal-insulator transition with critical interaction strength $U_c\approx 3D$.

\subsection{Doped}\label{sec::results::dopped}
\begin{figure}[t]
    \centering
    \includegraphics[width=1.00\columnwidth]{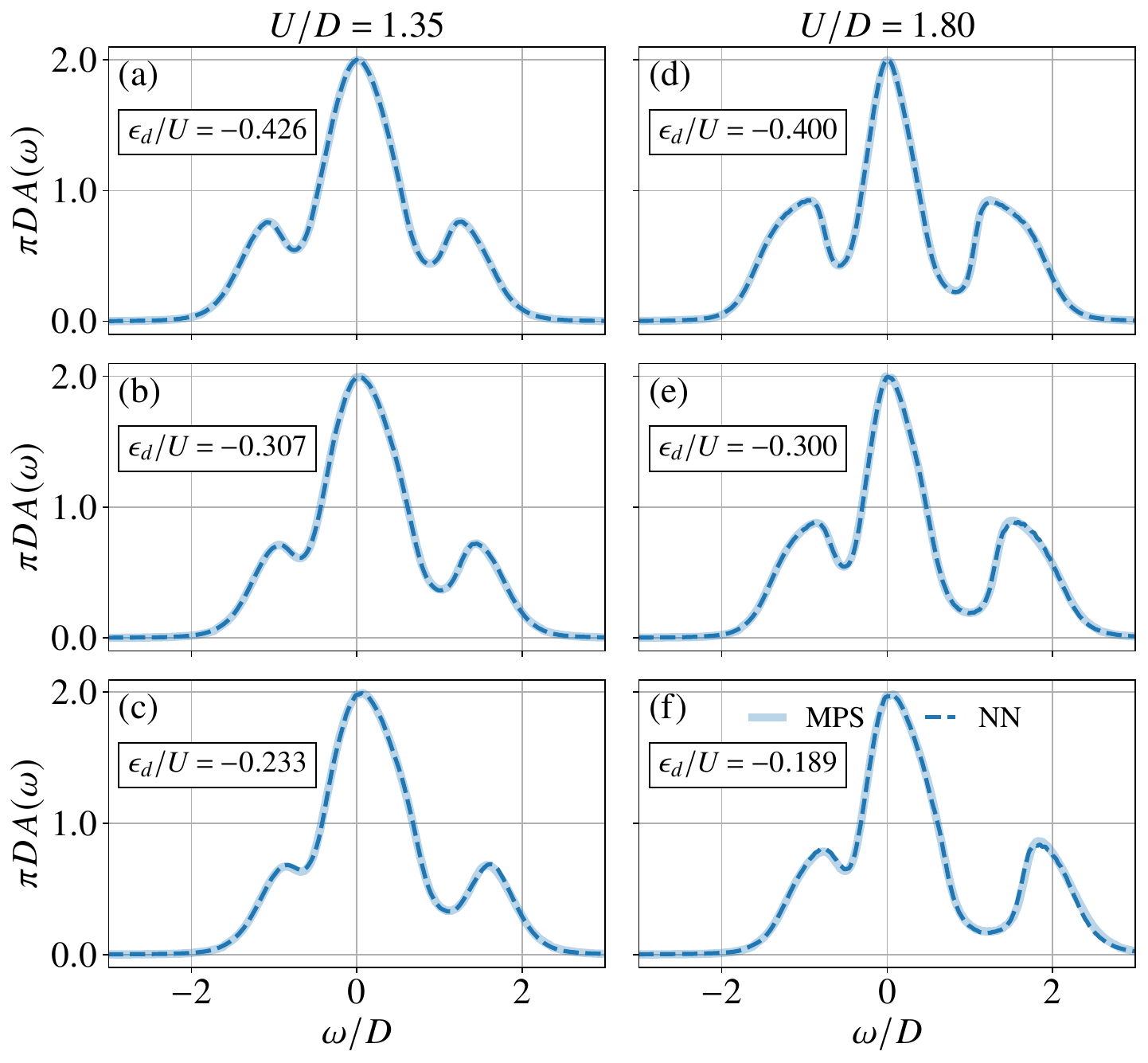}
    \caption{Spectral functions for hole-doped cases with $\langle \hat{n}_\sigma\rangle\sim 0.4$. 
    Panels (a)--(c) show results for $U/D = 1.35$, and panels (d)--(f) correspond to $U/D = 1.80$. 
    Thick lines represent reference results from the real-time MPS solver, and dashed lines denote predictions 
    from the neural network solver with $\gamma=1$.}
    \label{fig:dmftnn_nonhalf}
\end{figure}

\begin{figure}[t]
    \centering
    \includegraphics[width=1.00\columnwidth]{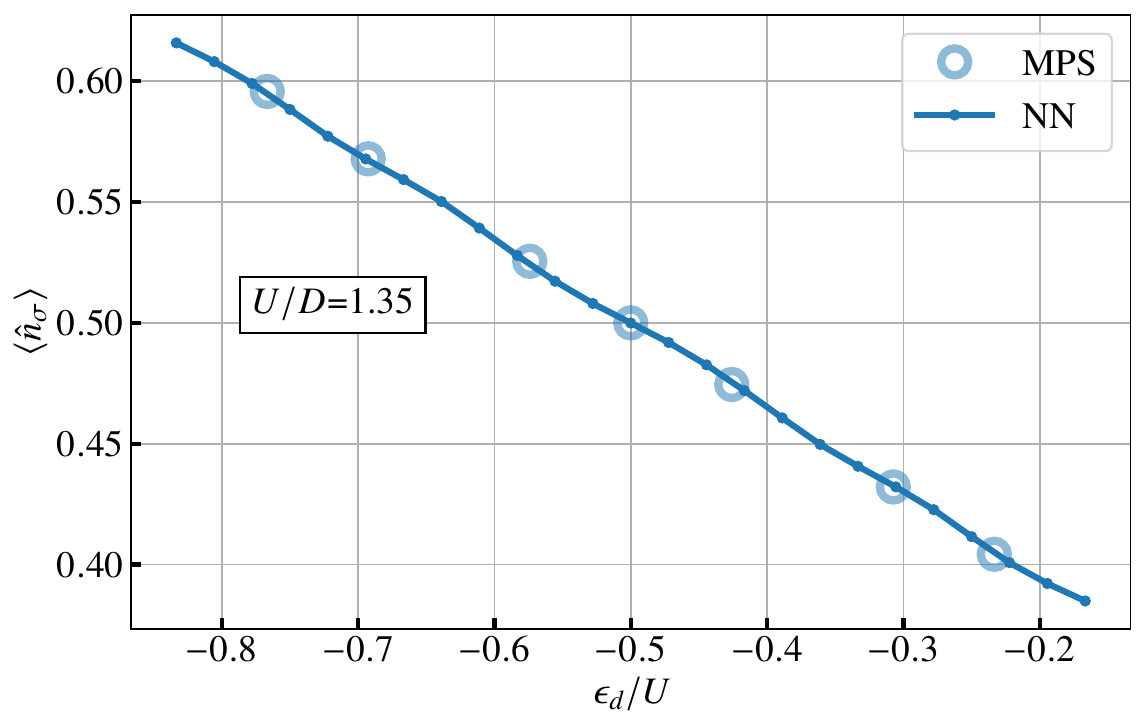}
    \caption{Filling $\langle \hat{n}_{\sigma}\rangle$ as a function of $\epsilon_d/U$ for the doped case at $U/D = 1.35$. }
    \label{fig:filldop}
\end{figure}

We further establish our solver's capability beyond half filling by demonstrating its performance across a wide doping range. As shown in Fig.~\ref{fig:dmftnn_nonhalf}, the solver accurately reproduces spectral functions for $U/D = 1.35, 1.80$ at a hole-doped filling of $\langle\hat{n}_\sigma\rangle \sim 0.4$. Moreover, Fig.~\ref{fig:filldop} confirms that the solver reliably predicts the filling level at $U/D = 1.35$ across different impurity energy levels $\epsilon_d$, showing excellent agreement with the reference data. These results collectively confirm the solver's reliability for full DMFT self-consistent calculations in the doped regime.

\subsection{Stability of DMFT self-consistency loop}

\begin{figure}[t]
    \centering
    \includegraphics[width=1.0\columnwidth]{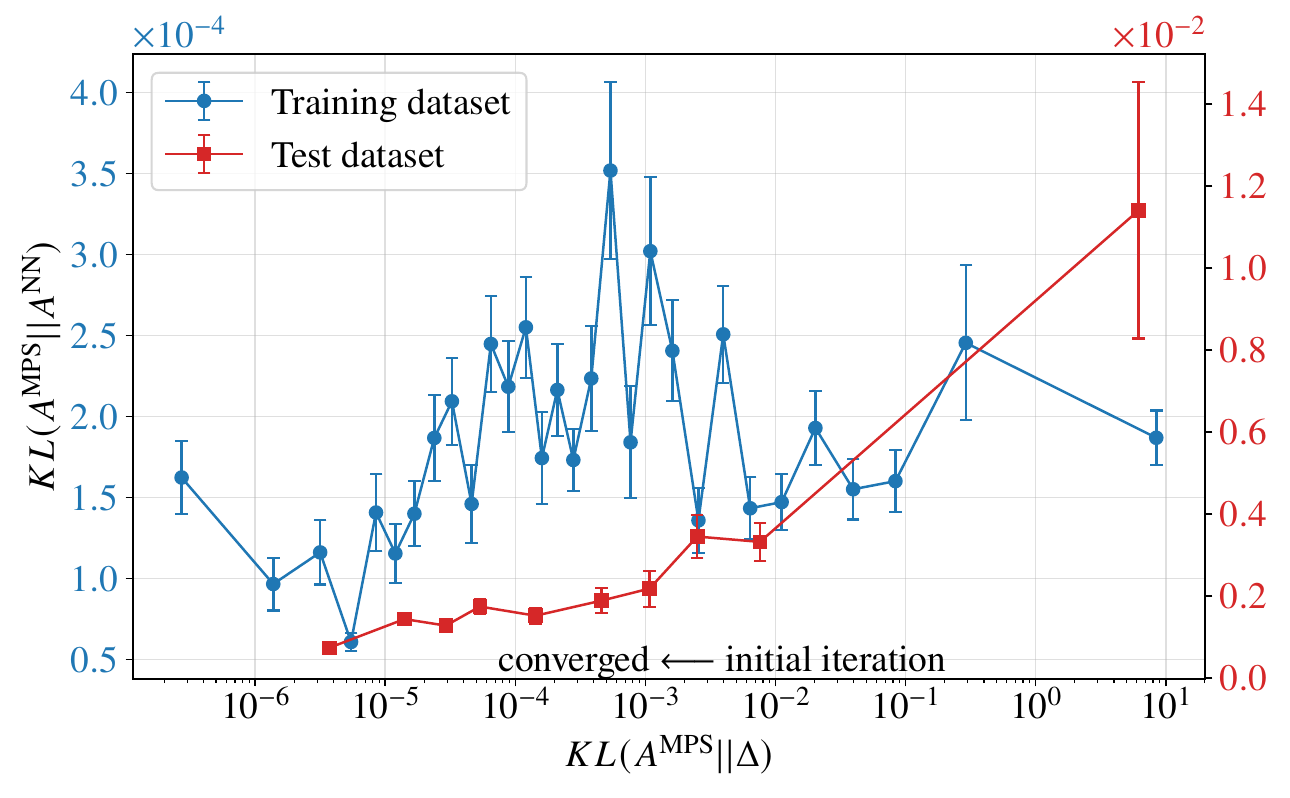}
    \caption{KL divergence between the MPS reference spectral function and the network prediction $\text{KL}(A^{\text{MPS}} \| A^{\text{NN}})$ plotted against the KL divergence between the MPS spectral function and the input hybridization $\text{KL}\bigl( A^{\text{MPS}}|| \Delta \bigr)$ for the training (blue, left axis) and test (red, right axis) datasets. The arrow indicates the direction of DMFT iterations from the initial step to the converged solution.
    }
    \label{fig:dkl}
\end{figure}

We assess the stability of our neural network impurity solver within the DMFT self-consistency loop by comparing prediction errors across different convergence stages. The error is quantified using the Kullback-Leibler (KL) divergence between two positive-definite spectral functions $\text{KL}\bigl(A^1||A^2\bigr) = \int d\omega A^1(\omega)\ln\frac{A^1(\omega)}{A^2(\omega)}$. 
For each sample, we compute the KL divergence between the MPS reference spectra and the input hybridization function, $\text{KL}\bigl( A^{\text{MPS}}|| \Delta \bigr)$, which serves as a proxy for the magnitude of the DMFT update: large values correspond to early iterations far from convergence, while small values characterize the converged regime.
We then examine the prediction error $\text{KL}(A^{\text{MPS}} \| A^{\text{NN}})$ as a function of $\text{KL}(A^{\text{MPS}} \| \Delta)$, as depicted in Fig.~\ref{fig:dkl}. On the training dataset, $\text{KL}(A^{\text{MPS}} \| A^{\text{NN}})$ remains consistently low over several orders of magnitude in $\text{KL}(A^{\text{MPS}} \| \Delta)$, indicating that the network has learned a stable mapping from the impurity inputs (hybridization and interactions) to the spectral function.
On the test dataset, we observe the expected trend that smaller input–output differences $\text{KL}\bigl( A^{\text{MPS}}|| \Delta \bigr)$ lead to systematically smaller prediction errors, whereas larger input–output differences correspond to larger, but still bounded errors $\text{KL}\bigl( A^{\text{MPS}}|| A^{\text{NN}} \bigr)$. This behavior ensures that as the DMFT loop approaches self-consistency, the neural network solver produces increasingly accurate updates and prevents error amplification from iteration to iteration. Taken together, these statistics demonstrate that our neural network impurity solver is numerically stable within the DMFT self-consistency loop.

\subsection{Interpretation of the latent space}\label{sec::results::latent}

\begin{figure*}[t]
    \centering
    \includegraphics[width=1.95\columnwidth]{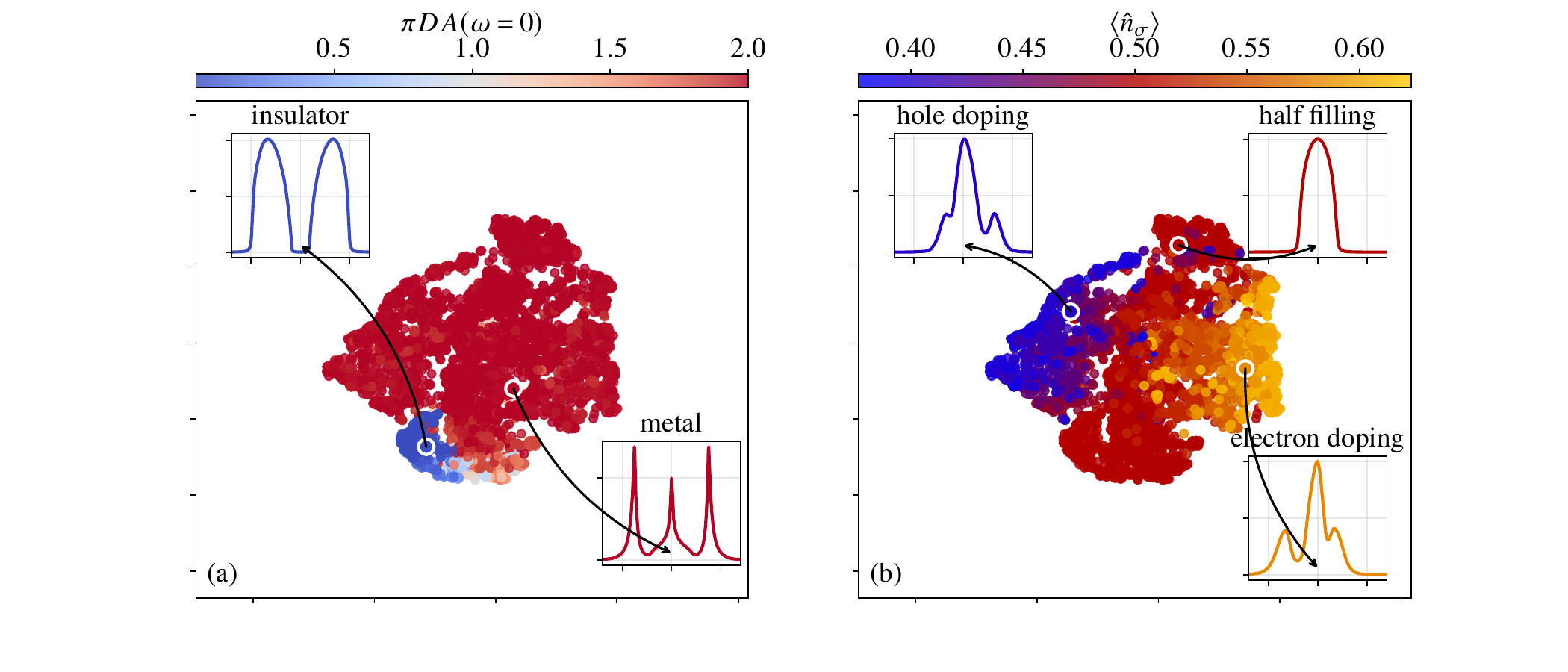}
    \caption{UMAP projections of the latent vectors collected from the cross-attention layers of the neural network for 5,441 samples. 
            (a) Samples are labeled by the spectral weight at the Fermi level $A_0 = A(\omega=0)$, separating metallic (nonzero $A_0$) and insulating ($A_0=0$) phases. 
            (b) Samples are labeled by the filling value, roughly clustering into half filling, hole doping, and electron doping cases.}
    \label{fig:umap}
\end{figure*}

To better understand the learned representations, we collect intermediate vectors from the cross-attention layers and analyze the network’s latent space. These vectors are first reduced to 50 dimensions using principal component analysis (PCA) and then projected to two dimensions with the Uniform Manifold Approximation and Projection (UMAP) algorithm \cite{mcinnes2020umap}. Each sample is labeled by two quantities: the spectral weight at the Fermi level $A(\omega=0)$ and the filling $\langle\hat{n}_{\sigma}\rangle$.
The resulting projections in Fig.~\ref{fig:umap} reveal a physically meaningful organization of the latent space. Panel (a) shows a clear separation into two clusters—metallic states with non-zero $A(\omega=0)$ and insulating states with vanishing spectral weight at the Fermi level. 
Meanwhile, panel (b) shows that samples group into three distinct regions corresponding to half filling, hole doping, and electron doping. This emergent clustering demonstrates that the network has internalized the essential physics, forming representations that reliably distinguish between metallic and insulating states as well as different doping regimes.

\section{Conclusions}\label{sec::conclusion}
This work introduces a neural network  impurity solver for single-orbital DMFT, designed as a fast and reliable alternative to conventional solvers. A key advantage is its direct output of real-frequency spectral functions. Trained on data from a complex-time solver with a derivative-informed loss function, the model learns a robust physical representation. Benchmark results confirm its accuracy in reproducing key physical features across metallic, insulating, and doped regimes.

Looking ahead, this framework can be extended to realistic lattice systems via real-space DMFT and to multi-band impurities with inter-orbital couplings and complex hybridization functions. Such extensions would provide fast, high-accuracy machine-learning solvers for strongly correlated systems beyond single-band models, thereby paving the way for practical real-frequency DMFT calculations in a broader class of materials.

\begin{acknowledgments}
X. C. acknowledges support from the National Key R\&D Program of China(No.2024YFA1408602). Y. L. acknowledges support from the National Key R\&D Program of China (No. 2022YFA1403000), the National Natural Science Foundation of China (No. 12274207), and the Basic Research Program of Jiangsu (Grant No. BK20253009). We greatly appreciate Xinyang Dong for the valuable feedback on the manuscript and for the insightful discussions throughout this project.
We thank Karsten Held for stimulating discussions.

\end{acknowledgments}

\appendix
\counterwithin{figure}{section}

\section{Computational details}\label{sec::method::comp_details}

\subsection{Solver parameters}
The training and test datasets are generated by performing DMFT calculations with the complex-time impurity solver. All calculations are performed with bond dimension $\chi=200$, time step $\mathrm{d}t=0.04/D$, and maximum evolution time $t_{\rm end}=100/D$. Spectra are evaluated over the frequency window $\omega \in [-5D,5D]$. DMRG and TDVP truncation weights are set to $10^{-15}$. 
Near the metal–insulator crossover, the metallic phase features a sharp Kondo resonance at low frequencies that finite real-time TDVP evolution cannot fully resolve. To obtain reliable training spectra, we apply linear prediction as a post-processing step to extend the effective time series and suppress low-frequency oscillations.

\subsection{Frequency discretization}
\begin{figure}[t]
    \centering
    \includegraphics[width=1.0\columnwidth]{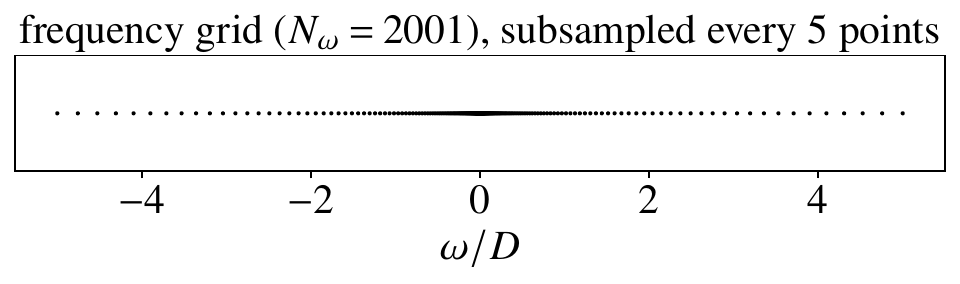}
    \caption{Illustration of the geometric frequency grid used for hybridization and spectral functions. The full grid contains $N_\omega = 2001$ points; for visualization purposes, only every fifth point is shown. The true spacing of consecutive intervals satisfies $\Delta\omega_{i+1}/\Delta\omega_i = 1.01$.}
    \label{fig:wgrid}
\end{figure}
The hybridization and spectral functions are discretized on a symmetric geometric grid with $N_\omega = 2001$ points as shown in Fig.~\ref{fig:wgrid}. 
The spacing of consecutive frequencies $\Delta\omega_i = |\omega_{i+1}-\omega_i|$ grows geometrically such that 
$\Delta\omega_{i+1}/\Delta\omega_i = 1.01$, yielding denser sampling near $\omega = 0$ and gradually coarsening at higher energies. 
This frequency discretization is inspired by the logarithmic discretization technique used in numerical renormalization group calculations~\cite{wilson1975renormalization, Bulla2008NRGReview}.

\subsection{Hyperparameters of the neural network}
We set the latent dimension to $d_\mathrm{emb}=2048$, which defines the size of the embeddings for both the hybridization function and the physical parameters. 
The physical parameters are embedded via a 5-layer FFN, while the hybridization function is embedded via a 4-layer FAN. 
Within each FAN layer the output channels are allocated to sine, cosine, and GELU activations in the ratio $1:1:2$. 

The embeddings are fused through two stacked cross-attention blocks of dimension $d_\mathrm{emb}$ with $H=32$ heads. After the first cross-attention and FFN layer, the resulting representation is treated as an updated hybridization function representation and fed into the second transformer block again, while the parameter embedding continues to act as query and key. Each attention block includes a small dropout of 2\% to avoid over-fitting.

\subsection{Training Details}
The network is trained using the Adam optimizer~\cite{kingma2014adam} with default $\beta$ parameters. 
The learning rate follows a cosine annealing schedule, which linearly ramps to $8\times10^{-5}$ and subsequently decays to $1\times10^{-6}$.

\bibliographystyle{apsrev4-2}
\bibliography{ref}

\end{document}